\documentstyle[12pt]{article}

\def\sfrac#1#2{{\textstyle{#1 \over #2}}}
\def\agt{\hbox{${\lower.40ex\hbox{$>$}
\atop \raise.20ex\hbox{$\sim$}}$}}
\def\alt{\hbox{${\lower.40ex\hbox{$<$}
\atop \raise.20ex\hbox{$\sim$}}$}}

\font\tenrm=cmr10
\font\tenit=cmti10
\font\elevenbf=cmbx10 scaled\magstep 1
\font\elevenrm=cmr10 scaled\magstep 1
\font\elevenit=cmti10 scaled\magstep 1

\font\ninerm=cmr9

\renewenvironment{thebibliography}[1]
 { \elevenrm
   \begin{list}{\arabic{enumi}.}
    {\usecounter{enumi} \setlength{\parsep}{0pt}
     \setlength{\itemsep}{3pt} \settowidth{\labelwidth}{#1.}
     \sloppy
    }}{\end{list}}

\begin{document}

\hfill SFU HEP-113-93

\begin{center}
\vglue 0.6cm
{{\elevenbf 
AN UPPER BOUND ON {\elevenit P}-WAVE CHARMONIUM
PRODUCTION VIA THE COLOR-OCTET MECHANISM}\footnote{
\ninerm
\baselineskip=11pt
Presented at the
{\it Workshop on Physics at Current Accelerators and the Supercollider}
at Argonne National Laboratory, June 1993.\\}
\vglue 1.0cm
{\tenrm HOWARD D. TROTTIER \\}
\baselineskip=13pt
{\tenit Simon Fraser University, Department of Physics, \\}
\baselineskip=12pt
{\tenit Burnaby, B.C. V5A 1S6 Canada\\}}

\vglue 0.8cm
{\tenrm ABSTRACT}

\end{center}

\vglue 0.3cm
{\rightskip=3pc
\leftskip=3pc
\tenrm\baselineskip=12pt
\noindent
A factorization theorem for $P$-wave quarkonium production,
recently derived by Bodwin, Braaten, Yuan and Lepage, is
applied to $\Upsilon \to \chi_{cJ} + X$, where $\chi_{cJ}$
labels the ${}^3 P_J$ charmonium states. The widths for
$\chi_{cJ}$ production through color-singlet $P$-wave 
and color-octet $S$-wave $c \bar c$ subprocesses are computed
each to leading order in $\alpha_s$. Experimental data on 
$\Upsilon \to J / \psi + X$ is used to obtain 
an upper bound on a nonperturbative parameter (related to the
probability for color-octet $S$-wave $c \bar c$ hadronization
into $P$-wave charmonium) that enters into the 
factorization theorem. 
}

\baselineskip=14pt

\vglue 0.6cm

Factorization theorems play a basic role in perturbative QCD
calculations of many hadronic processes. A well known factorization
theorem for the decay and production of $S$-wave quarkonium
follows from a nonrelativistic description of heavy quark-antiquark
($Q \overline Q$) binding \cite{Kwong}. Nonperturbative effects are 
factored into $R_S(0)$, the nonrelativistic wave function at the origin, 
leaving a hard $Q \overline Q$ subprocess matrix 
element that can be calculated in perturbation theory. 

Remarkably, the correct factorization theorems for 
the decay \cite{BBL} and production \cite{BBYL} of $P$-wave
quarkonium have only recently been derived. These new
theorems resolve a long standing problem regarding
infrared divergences which appear in some cases to leading order
in the rates for $P$-wave $Q \overline Q$ states \cite{Barbieri}.
In previous phenomenological calculations, the divergence was
replaced by a logarithm of a soft binding scale \cite{Barbieri,Kwong}.
However, a rigorous calculation requires that one consider
additional components of the Fock space for $P$-wave quarkonium,
such as $\vert Q \overline Q g \rangle$, where
the $Q \overline Q$ pair is in a color-octet $S$-wave state, 
and $g$ is a soft gluon \cite{BBL,BBYL}. 

A renewed study of the decay and production of $P$-wave quarkonium 
is therefore of considerable interest, since one may gain new
information on a nonperturbative sector of QCD that has largely
been neglected in the quark model description of heavy quarkonium.
This is also of practical consequence; for example, $J / \psi$ 
production provides a clean experimental signature for many
important processes, and $P$-wave charmonium states have 
appreciable branching fractions to $J / \psi$.

In this paper the factorization theorem for $P$-wave quarkonium
production is applied to $\Upsilon \to \chi_{cJ} + X$, where 
$\chi_{cJ}$ labels the ${}^3 P_J$ charmonium states. The widths 
for $\chi_{cJ}$ production through color-singlet $P$-wave 
and color-octet $S$-wave $c \bar c$ subprocesses are computed
each to leading order in $\alpha_s$. Experimental data on 
$\Upsilon \to J / \psi + X$ is used to obtain 
an upper bound on a nonperturbative parameter (related to the
probability for color-octet $S$-wave $c \bar c$ hadronization
into $P$-wave charmonium) that enters into the 
factorization theorem. 
The bound obtained here adds to the limited information so 
far available on the color-octet mechanism for $P$-wave 
quarkonium production.
The color-octet component in $P$-wave 
decay was estimated in Ref. \cite{BBL} from measured
decay rates of the $\chi_{c1}$ and $\chi_{c2}$. A rough estimate
of the color-octet component in $P$-wave charmonium production 
was obtained in Ref. \cite{BBYL} from data on $B$ meson
decays; however, an accurate determination in that case
requires a calculation of next-to-leading order QCD corrections 
to the color-singlet component of $B \to \chi_{cJ} + X$,
which is so far unavailable \cite{BBYL}.

The factorization theorem for $P$-wave quarkonium production
has two terms, and in the case of $\Upsilon$ decay takes the form:
\begin{eqnarray}
   \Gamma(\Upsilon \to \chi_{cJ} + X)
   & = & H_1 \hat 
   \Gamma_1(\Upsilon \to c \bar c ( {}^3 P_J ) + X ; \mu)
\nonumber \\ 
   & & \mbox{} + (2 J + 1) H_8'(\mu) 
   \hat \Gamma_8(\Upsilon \to c \bar c ( {}^3 S_1 ) + X) .
\label{Fact}
\end{eqnarray}
$\hat \Gamma_1$ and $\hat \Gamma_8$ are hard subprocess rates
for the production of a $c \bar c$ pair in color-singlet $P$-wave 
and color-octet $S$-wave states respectively. The quarks 
are taken to have vanishing relative momentum. 
The nonperturbative parameters $H_1$ and $H_8'$ are proportional 
to the probabilities for these $c \bar c$ configurations to hadronize 
into a color-singlet $P$-wave bound state. $H_1$, 
$H_8'$ and $\hat \Gamma_8$ are independent of $J$.
This factorization theorem is valid to all orders in $\alpha_s$
and to leading order in $v^2$, where $v$ is the typical 
center-of-mass velocity of the heavy quarks.
The hard subprocess rates are free 
of infrared divergences. $\hat \Gamma_1$ and $H'_8$ depend on an 
arbitrary factorization scale $\mu$ in such a way that the physical
decay rate is independent of $\mu$. In order to avoid large
logarithms of $m_\Upsilon / \mu$ in $\hat \Gamma_1$,
$\mu$ of $O(m_\Upsilon)$ should be used.

$H_1$ can be expressed 
in terms of the $P$-wave color-singlet $c \overline c$  
wave function,
$H_1 \approx 9 \vert R_P'(0) \vert^2 / (2\pi m_c^4) 
     \approx 15~{\rm MeV}$,
where the numerical estimate was obtained in Ref. \cite{BBL}
from measured decay rates of the $\chi_{c1}$ and $\chi_{c2}$.
$H_8'$ cannot be rigorously expressed perturbatively in terms 
of $R_P$, since it accounts for radiation of a soft gluon by 
a color-octet $c \bar c$ pair. 
The scale dependence of $H_8'(\mu)$ is determined by a
renormalization-group equation \cite{BBL,BBL2}
which (at leading order in $\alpha_s(\mu)$) gives \cite{BBYL}:
\begin{equation}
   H_8'(m_b) = H_8'(\mu_0) + \left[ 
   {16 \over 27 \beta_3} 
   \ln\left( {\alpha_s(\mu_0) \over \alpha_s(m_c)} \right)
 + {16 \over 27 \beta_4}
   \ln\left( {\alpha_s(m_c) \over \alpha_s(m_b)} \right)
   \right] H_1 
\label{H8Mb}
\end{equation}
(for $\mu_0 < m_c$), where $\beta_n = (33 - 2n) / 6$.
If $H_8'(\mu_0)$ is neglected in the limit of large $m_b$
one obtains $H_8'(m_b) \approx 3$~MeV,
using $\alpha_s(\mu_0) \sim 1$ \cite{BBYL}. While one might
not expect the physical value of $m_b$ to be large enough 
to neglect $H_8'(\mu_0)$, an estimate for $H_8'(m_b)$ 
obtained in Ref. \cite{BBYL} from experimental data on 
$B$ meson decays is consistent with the above result. 

A calculation of $\hat \Gamma_1$ and $\hat \Gamma_8$
in Eq. (\ref{Fact}) each to leading order in $\alpha_s$ 
can be obtained from a calculation of the infrared divergent width 
$\Gamma_{\rm div}$ for $\Upsilon \to c \bar c ({}^3 P_J) + ggg$,
where the $c \bar c$ pair is in a color-singlet $P$-wave state:
\begin{eqnarray}
\lefteqn{
   \Gamma_{\rm div}
   (\Upsilon \to c \bar c ({}^3 P_J) + ggg; \mu_0) 
   \equiv
        }
\nonumber \\ 
   & & {20 \alpha_s^5 \over 3^7 \pi^3} 
   {G^\Upsilon_1 \over m_\chi}
   \left[ {\cal F}_{1J}(\mu) + 
   (2J + 1) {16 \over 27\pi} 
   \ln\left( {\mu \over \mu_0} \right) {\cal F}_8 \right] H_1 . 
\label{Gdiv}
\end{eqnarray}
${\cal F}_{1J}$ and ${\cal F}_8$ are dimensionless infrared-finite 
form factors. $\mu_0$ is an infrared cutoff on the energy of 
soft gluons, and $\mu$ is an arbitrary factorization scale
[the $\mu$ dependence of ${\cal F}_{1J}$ exactly cancels that
of the explicit logarithm in Eq. (\ref{Gdiv})].
The constants in Eq. (\ref{Gdiv}) include a color-factor of $5 / 216$ 
and phase space factors, including $1/3$ for $\Upsilon$ spin-averaging,
and $1/3!$ for the phase space of the three indistinguishable gluons
[cf. Eq. (\ref{Phase}) below].
$G^\Upsilon_1$ is related to the usual $S$-wave $b \bar b$ 
nonrelativistic wave function,
$G^\Upsilon_1 \approx 
   3 \vert R^\Upsilon_S(0) \vert^2 / (2\pi m_b^2)
   \approx 108~{\rm MeV}$,
where the numerical value is obtained from the electronic decay
rate of the $\Upsilon$ \cite{PDG}.

The hard subprocess rates of Eq. (\ref{Fact}) are identified
from $\Gamma_{\rm div}$ by using the perturbative expression
for the infrared divergence in $H_8'$, obtained by
neglecting the running of the coupling \cite{BBYL}:
$H_8'(\mu) \sim (16 / 27\pi) \alpha_s \ln(\mu / \mu_0) H_1$.
Thus:
\begin{equation}
   \hat \Gamma_1(\Upsilon \to c \bar c ({}^3 P_J) + ggg; \mu)
   = {20 \alpha_s^5 \over 3^7 \pi^3} 
     {G^\Upsilon_1 \over m_\chi} 
     {\cal F}_{1J}(\mu) ,
\label{hat1}
\end{equation}
and
\begin{equation}
   \hat \Gamma_8(\Upsilon \to c \bar c ({}^3 S_1) + gg)
   = {20 \alpha_s^4 \over 3^7 \pi^3} 
     {G^\Upsilon_1 \over m_\chi} 
     {\cal F}_8 .
\label{hat8}
\end{equation}
Note that $\hat\Gamma_1$ is suppressed by $O(\alpha_s)$
compared to $\hat\Gamma_8$. However, the nonperturbative 
parameters $H_1$ and $H_8'$ which accompany these subprocess
rates in Eq. (\ref{Fact}) are independent, hence $\alpha_s H_1$
need not be small compared to $H_8'$ \cite{BBL,BBYL}.
We therefore proceed to calculate $\hat\Gamma_1$
and $\hat\Gamma_8$ each to leading order; all further 
corrections to $P$-wave production are then guaranteed
to be suppressed by at least one power of $\alpha_s$
compared to what is included here. 

In order to extract ${\cal F}_{1J}$ and ${\cal F}_8$ individually,
it is necessary to explicitly identify the infrared logarithm in 
the calculation of $\Gamma_{\rm div}$. This can be done 
analytically, as described in the following.
\begin{figure}
\vspace{3in}
\caption{\tenrm\baselineskip=12pt
One of the 36 $O(\alpha_s^5$) diagrams contributing
to $\Upsilon \to c \bar c ({}^3 P_J) + ggg$.\label{FigFeynman}}
\end{figure}
There are 36 $O(\alpha_s^5)$ diagrams contributing to 
$\Gamma_{\rm div}$. One of these is shown in 
Fig. \ref{FigFeynman}. 
Define the invariant amplitude ${\cal M}_J(2,3;1)$
corresponding to the sum of all Feynman diagrams
where gluon ``1'' is radiated from the charm quark line.
The amplitude is readily computed using expressions 
for $S$- and $P$-wave $Q \overline Q$ currents given in 
Ref. \cite{Kuhn}%
\footnote{\ninerm\baselineskip=11pt Overall factors in the quark currents
including couplings, color amplitudes, and wave functions
have been accounted for in Eq. (\ref{Gdiv}).}
\begin{equation}
   {\cal M}_J(2,3;1) \equiv -
   { m_\Upsilon m_\chi  B_\mu(2,3)  C_J^\mu(1)
   \over 
   [ (k_2 + k_4) \cdot k_3 ] [ (k_3 + k_4) \cdot k_2 ]
   [ (k_2 + k_3) \cdot k_4 ] \, k_4^2 \, (k \cdot k_1)^2 } ,
\label{M231}
\end{equation}
where 
\begin{eqnarray}
   \epsilon_4^\mu B_\mu(2,3) & = &
   \bigl\{ 
   \epsilon_4 \cdot \epsilon_2 [ 
     - k_4 \cdot k_3 \epsilon_3 \cdot k_2 \epsilon_0 \cdot k_4
     - k_2 \cdot k_3 \epsilon_3 \cdot k_4 \epsilon_0 \cdot k_2
     - k_4 \cdot k_3 k_2 \cdot k_3 \epsilon_0 \cdot \epsilon_3 ]
\nonumber \\ 
   & & \mbox{} + 
   \epsilon_0 \cdot \epsilon_3 [
       k_4 \cdot k_3 \epsilon_4 \cdot k_2 \epsilon_2 \cdot k_3
     + k_2 \cdot k_3 \epsilon_2 \cdot k_4 \epsilon_4 \cdot k_3
     - k_4 \cdot k_2 \epsilon_4 \cdot k_3 \epsilon_2 \cdot k_3 ]
   \bigr\}
\nonumber \\ 
   & & \mbox{} + \left\{ 2 \leftrightarrow 3 \right\}
               + \left\{ 3 \leftrightarrow 4 \right\} 
\label{B23}
\end{eqnarray}
($\epsilon_0$ is the polarization of the $\Upsilon$), and
\begin{eqnarray}
   & & \epsilon_{4\mu} C_{J=0}^\mu(1) =
   \sqrt{\sfrac16} 
   \left[ \epsilon_1 \cdot \epsilon_4 k_1 \cdot k_4
        - \epsilon_1 \cdot k_4 \epsilon_4 \cdot k_1 \right]
   \left( m_\chi^2 + k \cdot k_4 - k_4^2 \right) ,
\nonumber \\ 
   & & \epsilon_{4\mu} C_{J=1}^\mu(1) =
   \sfrac12 m_\chi k_4^2 \,
   \varepsilon_{\alpha\beta\gamma\delta} \,
   e^\alpha \epsilon_4^\beta \epsilon_1^\gamma k_1^\delta ,
\label{C1} \\ 
   & & \epsilon_{4\mu} C_{J=2}^\mu(1) =
   \sqrt{\sfrac12} m_\chi^2 \left(
   k_1 \cdot k_4 \epsilon_1^\alpha \epsilon_4^\beta  +  
   k_4^\alpha k_1^\beta \epsilon_1 \cdot \epsilon_4  -
   k_1^\alpha \epsilon_4^\beta \epsilon_1 \cdot k_4  -
   k_4^\alpha \epsilon_1^\beta \epsilon_4 \cdot k_1 \right) 
   e^{\alpha\beta} .
\nonumber
\end{eqnarray}
$e^\alpha$ is a spin-1 polarization vector 
and $e^{\alpha\beta}$ is a spin-2 polarization tensor.
For convenience the virtual gluon is labeled in 
Eqs. (\ref{M231})--(\ref{C1}) by polarization $\epsilon_4$ 
and momentum $k_4$ ($k_4 = P - k_2 - k_3 = k + k_1$).
Terms which vanish due to the on-shell conditions
$\epsilon_i \cdot k_i = 0$ ($i=1,2,3$) and 
$\epsilon_0 \cdot P = 0$ have been dropped.

The overall factors in Eq. (\ref{Gdiv}) are such that:
\begin{eqnarray}
\lefteqn{
   {\cal F}_{1J}(\mu) + 
   (2J + 1) {16 \over 27\pi} 
   \ln\left( {\mu \over \mu_0} \right) {\cal F}_8
   \equiv 
        }
\nonumber \\ 
   & & 3 \int d[\Phi_4] \sum_{\rm spins} 
   \left[ {\cal M}_J^2(2,3;1) 
      + 2 {\cal M}_J(2,3;1) {\cal M}_J(1,3;2) \right] ,
\label{Phase}
\end{eqnarray}
where $\Phi_n$ denotes (infrared-cutoff) $n$-body phase space, 
normalized according to
\begin{equation}
   \Phi_n[P \to p_1,\ldots,p_n] \equiv
   \int \prod_{i=1}^n {d^3p_i \over 2E_i}
   \delta^4(P - \sum_i p_i) .
\label{Phi}
\end{equation}
The factor of 3 on the right hand side of Eq. (\ref{Phase}) 
accounts for symmetrization of ${\cal M}_J(2,3;1)$
under gluon label interchanges $1 \leftrightarrow 2$
and $1 \leftrightarrow 3$, taking account
of the symmetry in the three gluon phase space.

The infrared divergence comes entirely from the first 
term in square brackets in Eq. (\ref{Phase}), 
and is due to the $P$-wave
charm quark propagator $1/(k \cdot k_1)^2$ in Eq. (\ref{M231}).
It is therefore advantageous to organize the four-body
phase space integral in Eq. (\ref{Phase}) by taking 
the invariant mass of the $\chi_{cJ}$ and gluon ``1'' as one
integration variable \cite{Byckling}
\begin{eqnarray}
\lefteqn{
   \int d[\Phi_4] = 
   \int_0^{(m_\Upsilon - m_\chi)^2} 
        d(k_{23}^2) \,
   \int_{m_\chi^2 + 2 \mu_0 m_\chi}^{(m_\Upsilon - m_{23})^2} 
        d(k_{1\chi}^2) \,
        }
\nonumber \\
   & & \mbox{} \times
       \Phi_2 [ P \to k_{23} , k_{1\chi} ] \,
       \Phi_2 [ k_{23} \to k_2 , k_3 ] \,
       \Phi_2 [ k_{1\chi} \to k_1 , k ] ,
\label{Phi4}
\end{eqnarray}
where $m_{23}^2 \equiv k_{23}^2$.
Note the infrared cutoff $\mu_0$ on the energy of gluon ``1''
in the rest frame of the $\chi_{cJ}$.

The infrared logarithm on the right-hand side of Eq. (\ref{Phase})
can now be identified analytically by observing that
$B_\mu(2,3) C^\mu_J(1)$ in Eq. (\ref{M231}) is given by a 
sum of terms each containing exactly one factor of $k_1$, if
$k_4 = P - k_2 - k_3$ is used to eliminate the virtual gluon 
momentum. With this convention, one has
\begin{equation}
   \sum_{\rm spins} {\cal M}_J^2(2,3;1) =
   {\gamma_J(k_1; P, k, k_2, k_3) \over (k \cdot k_1)^2} ,
\label{gk1}
\end{equation}
where $k_1$ appears explicitly in the function 
$\gamma_J(k_1; P, k, k_2, k_3)$ only in the combination
$k_1 / k \cdot k_1$. 
${\cal F}_8$ is then given in terms of a manifestly 
infrared-finite three-body phase space integral, taking account 
of the fact that 
$\Phi_2( k_{1\chi} \to k_1 , k ) 
= \sfrac14 k \cdot k_1 / k_{1\chi}^2 \int d\Omega_{1\chi}^*$,
where $\Omega_{1\chi}^*$ is the center-of-mass solid
angle of the two body system:
\begin{eqnarray}
   (2J + 1) {\cal F}_8 & = & 
   {27 \pi \over 32 m_\chi^2}
   \int_0^{(m_\Upsilon - m_\chi)^2} 
       d(k_{23}^2) \,
      \Phi_2 [ P \to k_{23} , k ]
\nonumber \\ 
   & & \mbox{} \times 
      \Phi_2 [ k_{23} \to k_2 , k_3 ]
   \int d\Omega_{1\chi}^* \, 
      \gamma_J(\widetilde k_1; P, k, k_2, k_3) ,
\label{F8int}
\end{eqnarray}
where
\begin{equation}
   \widetilde k_1 \equiv \lim_{k \cdot k_1 \to 0} 
   {k_1 \over k \cdot k_1} .
\label{k1soft}
\end{equation}
The finite four-vector $\widetilde k_1$ is readily expressed 
directly in terms of $k_{23}^2$ and $\Omega_{1\chi}^*$.
An expression for ${\cal F}_{1J}$ can be obtained 
from Eqs. (\ref{Phase}) and (\ref{F8int}) by analogy with
the identity
$\int dx f(x) / x = f(0) \ln x + \int dx [f(x) - f(0)] / x$.

The contraction of currents and sum over polarizations 
in Eqs. (\ref{M231}) and (\ref{Phase}) were performed 
symbolically using {\small REDUCE} \cite{REDUCE} 
(leading to lengthy expressions, particularly for $J=2$). 
The $\chi_{cJ}$ spin sums were done using 
(see e.g. Ref. \cite{Kuhn}):
\begin{eqnarray}
   & & \sum_e e_\mu e_\nu = - g_{\mu\nu} 
    + {k_\mu k_\nu \over m_\chi^2} \equiv {\cal P}_{\mu\nu} ,
\nonumber \\ 
   & & \sum_e e_{\mu\nu} e_{\alpha\beta}
    = \sfrac12 \left[ 
      {\cal P}_{\mu\alpha} {\cal P}_{\nu\beta}
    + {\cal P}_{\mu\beta}  {\cal P}_{\nu\alpha} \right]
    - \sfrac13
      {\cal P}_{\mu\nu} {\cal P}_{\alpha\beta} .
\label{chipol}
\end{eqnarray}
The phase space integrals were evaluated numerically using 
{\small VEGAS} \cite{VEGAS}; modest integration grids are 
found to give very good convergence.
The fact that ${\cal F}_8$ should be independent of $J$
provides a stringent check of these calculations, given that
the three currents $C_J^\mu$ have very different structures
[cf. Eq. (\ref{C1})]. This was verified explicitly in 
numerical calculations of Eq. (\ref{F8int}), to better
than a few tenths of a percent for all $m_\chi / m_\Upsilon$
on a modest integration grid.
Figure \ref{FigF8} shows the numerical results for ${\cal F}_8$
over a range of hypothetical meson masses.
In Fig. \ref{FigF1} results for ${\cal F}_{1J}(\mu)$ are shown 
using a factorization scale $\mu = m_\Upsilon$.
\begin{figure}
\vspace{3in}
\caption{\tenrm\baselineskip=12pt
Color-octet form factor ${\cal F}_8$ as a function
of $m_\chi / m_\Upsilon$.\label{FigF8}}
\end{figure}
\begin{figure}
\vspace{3in}
\caption{\tenrm\baselineskip=12pt
Color-singlet form factors ${\cal F}_{1J}$ as functions
of $m_\chi / m_\Upsilon$: $J=0$ (short-dashed line),
$J=1$ (long-dashed line), $J=2$ (solid line). The
form-factors were evaluated using a factorization scale
$\mu = m_\Upsilon$.\label{FigF1}}
\end{figure}

The available experimental data on charmonium production
in $\Upsilon$ decay is for the $J / \psi$:
\begin{equation}
   B_{\rm exp} (\Upsilon \to J / \psi + X)
   \ \left\{
   \begin{array}{ll}
   = (1.1 \pm 0.4) \times 10^{-3} 
       & \mbox{{\small CLEO} \cite{CLEO},}    \\
   <  1.7 \times 10^{-3}
       & \mbox{Crystal Ball \cite{Crystal},}  \\ 
   <  0.68 \times 10^{-3}
       & \mbox{{\small ARGUS} \cite{ARGUS}.}
   \end{array}
   \right.
\label{Bexp}
\end{equation}
An upper bound on $H_8'$ can be extracted from this data
by computing the ``indirect'' production 
of $J / \psi$ due to the $\chi_{cJ}$ states.
Assuming that radiative cascades from $\chi_{c1}$ 
and $\chi_{c2}$ dominate, with branching fractions
$B_{\rm exp}(\chi_{c1} \to \gamma J / \psi) \approx 27\%$
and
$B_{\rm exp}(\chi_{c2} \to \gamma J / \psi) \approx 13\%$
\cite{PDG}, the results presented here give:
\begin{equation}
   H_8'(m_\Upsilon) \approx \left\{ 
   { \sum_J B(\Upsilon \to \chi_{cJ} + X' \to J / \psi + X) 
     \over 2.9 \times 10^{-5} }
   + 1.4 \right\} \mbox{MeV} .
\label{H8Bpsi}
\end{equation}
The first number in brackets above comes from the
color-octet subprocess rate $\hat \Gamma_8$, and the
second number from the color-singlet rate $\hat \Gamma_1$.
The experimental value for the total width
$\Gamma_{\rm tot}(\Upsilon) \approx 52$~keV \cite{PDG}
was used, along with 
$\alpha_s(m_\Upsilon) \approx 0.179$ \cite{Kwong},
and the values of $H_1$ and $G^\Upsilon_1$ given above.

Equation (\ref{H8Bpsi}) yields the bound 
$H_8'(m_\Upsilon) \alt 25$~MeV
using the {\small ARGUS} upper limit, which is consistent 
with the other measurements. This bound is considerably larger
than an estimate $H_8'(m_b) \approx 3$~MeV
based on $B$ meson decays \cite{BBYL},%
\footnote{\ninerm\baselineskip=11pt
$H_8'(\mu)$ increases by only $\approx 0.3$~MeV in the 
evolution from $\mu=m_b$ to $\mu=m_\Upsilon$
[cf. Eq. (\ref{H8Mb})].}
although a calculation of next-to-leading order 
QCD corrections to the color-singlet component of
$B \to \chi_{cJ} + X$ is required before an accurate
determination of $H_8'$ can be made in that case \cite{BBYL}.

This raises the possibility of significant direct production 
of $J / \psi$ in the decay of the $\Upsilon$, 
unless the branching fraction turns out to be
considerably smaller than the {\small ARGUS} bound. Mechanisms 
for direct $\Upsilon \to J / \psi + X$ in perturbative QCD were first 
discussed in Refs. \cite{Fritzsch} and \cite{Bigi}. The direct
production rate is suppressed by $O(\alpha_s^2)$ compared to the
$P$-wave color-octet production mechanism considered here.
However, the nonperturbative matrix elements which enter into 
$P$-wave production are of $O(v^2)$ relative to the corresponding 
parameter for $S$-wave production.

The full $O(\alpha_s^6)$ perturbative QCD amplitude for 
direct $\Upsilon \to J / \psi + X$ was recently evaluated
in Ref. \cite{Irwin}, corresponding to one-loop diagrams
for $\Upsilon \to J / \psi + gg$, and tree diagrams
for $\Upsilon \to J / \psi + gggg$. 
The $O(\alpha_s^2 \alpha^2)$ electromagnetic amplitude
for the two gluon decay mode was also evaluated.
Unfortunately, only a crude estimate of the required phase space 
integrations was made in Ref. \cite{Irwin} (there is a costly 
convolution with a numerical calculation of the loop 
integrals for $\Upsilon \to J / \psi + gg$).
Nevertheless, the calculation of Ref. \cite{Irwin}
suggests a branching fraction for direct production
of a $\mbox{few} \times 10^{-4}$.
This would lead to a considerable reduction in the bound
on $H_8'$ extracted from Eqs. (\ref{Bexp}) and (\ref{H8Bpsi}).

To summarize, a complete calculation was made of the leading order
rates for $\Upsilon \to \chi_{cJ} + X$, through both color-singlet 
$P$-wave and color-octet $S$-wave $c \bar c$ subprocesses.
Experimental data on $J / \psi$ production 
was used to obtain an upper bound on the nonperturbative 
parameter $H_8'$, related to the probability for
color-octet $S$-wave $c \bar c$ hadronization into 
$P$-wave charmonium. 
Improved experimental data, and a
definitive calculation of the direct $J / \psi$ production rate 
along the lines of Ref. \cite{Irwin}, would allow for an
accurate determination of $H_8'$ from the results presented here. 

I am indebted to Eric Braaten for suggesting this problem,
and for many enlightening conversations. I also thank 
Mike Doncheski, John Ng, and Blake Irwin for helpful 
discussions. This work was supported in part by the 
Natural Sciences and Engineering Research Council of Canada.

\vglue 0.6cm


\begin{thebibliography}{99}

\bibitem{Kwong} See, for example, W. Kwong et al.,
{\elevenit Phys. Rev.} {\elevenbf D37} (1988) 3210.

\bibitem{BBL} G. T. Bodwin, E. Braaten, and G. P. Lepage,
{\elevenit Phys. Rev.} {\elevenbf D46} (1992) R1914.

\bibitem{BBYL} G. T. Bodwin, E. Braaten, T. C. Yuan,
and G. P. Lepage, {\elevenit Phys. Rev.} 
{\elevenbf D46} (1992) R3703.

\bibitem{Barbieri} R. Barbieri, R. Gatto, and E. Remiddi,
{\elevenit Phys. Lett.} {\elevenbf B61} (1976) 465;
R. Barbieri, M. Caffo, and E. Remiddi, 
{\elevenit Nucl. Phys.} {\elevenbf B162} (1980) 220;
R. Barbieri et al.,
{\elevenit Phys. Lett.} {\elevenbf B95} (1980) 93; 
{\elevenit Nucl. Phys.} {\elevenbf B192} (1981) 61.

\bibitem{BBL2} G. T. Bodwin, E. Braaten, G. P. Lepage
(in progress).

\bibitem{PDG} Particle Data Group, K. Hikasa et al.,
{\elevenit Phys. Rev.} {\elevenbf D45} (1992) S1.

\bibitem{Kuhn} J. H. K\"uhn, J. Kaplan, and E. G. O. Safiani,
{\elevenit Nucl. Phys.} {\elevenbf B157} (1979) 125;
J. G. K\"orner et al.,
{\it ibid.} {\elevenbf B229} (1983) 115.

\bibitem{Byckling} See, for example, E. Byckling and K. Kajantie,
{\elevenit Particle Kinematics\/} (John Wiley \& Sons, London, 1973).

\bibitem{REDUCE} A. C. Hearn, REDUCE {\elevenit User's Manual\/}
(Rand Corp. Publication CP78, 1984).

\bibitem{VEGAS} G. P. Lepage,
Cornell University Report No. CLNS-80/447 (1980).

\bibitem{CLEO} CLEO Collaboration, R. Fulton et al.,
{\elevenit Phys. Lett.} {\elevenbf B224} (1989) 445.

\bibitem{Crystal} Crystal Ball Collaboration, W. Maschmann et al.,
{\elevenit Z. Phys.} {\elevenbf C46} (1990) 555.

\bibitem{ARGUS} ARGUS Collaboration, H. Albrecht et al.,
{\elevenit Z. Phys.} {\elevenbf C55} (1992) 25.

\bibitem{Fritzsch} H. Fritzsch and K.-H. Streng,
{\elevenit Phys. Lett.} {\elevenbf B77} (1978) 299.

\bibitem{Bigi} I. I. Y. Bigi and S. Nussinov,
{\elevenit Phys. Lett.} {\elevenbf B82} (1979) 281.

\bibitem{Irwin} B. Irwin, Ph.D. thesis, McGill University (1991).

\end{thebibliography}
\end{document}